\documentclass{aa}
\usepackage{amsmath}

\usepackage[svgnames]{xcolor}
\definecolor{blue}{RGB}{0,0,255}
\usepackage{caption}
\usepackage[font={color=blue},figurename=Fig.,labelfont={it}]{caption}
\usepackage{bm}
\usepackage{subcaption}
\usepackage[varg]{txfonts}
\usepackage{graphicx}
\usepackage{natbib}
\usepackage{pdfpages}
\usepackage{wasysym}
\usepackage{rotating}
\bibpunct{(}{)}{;}{a}{}{,} 
\usepackage{hyperref}
\usepackage{indentfirst}

\hypersetup{
    colorlinks=true,                          
    linkcolor=blue, 
    citecolor=blue, 
    urlcolor=blue,
    linktoc=all
}

\begin{document}
\title{Triskels and Symmetries of Mean Global Sea-Level Pressure}
\author{F. Lopes\inst{1}
\and V. Courtillot\inst{1}
\and J-L. Le Mouël\inst{1}
}
\institute{{ Universit\'e de Paris, Institut de Physique du globe de Paris, CNRS UMR 7154, F-75005 Paris, France}
}
\date{}

\abstract {The evolution of mean sea-level atmospheric pressure since 1850 is analyzed using singular spectrum analysis. Maps of the main components (the trends) reveal striking symmetries of order
 3 and 4. The northern hemisphere (\textbf{NH}) displays a set of three positive features, forming an almost perfect equilateral triangle. The southern hemisphere (\textbf{SH}) displays a set of three positive features  arranged as an isosceles triangle, with a possible fourth (weaker) feature. This geometry can be modeled as Taylor-Couette flow of mode 3 (\textbf{NH}) or 4 (\textbf{SH}). The remarkable regularity and order  three symmetry of the \textbf{NH} triskel occurs despite the lack of cylindrical symmetry of the northern continents. The stronger intensity and larger size of features in the \textbf{SH} is linked to the presence of  the annular \textbf{AAO}. In addition to the dominant trends, quasi-periodic components of $\sim$130, 90, 50,  22, 15, 4, 1.8, 1, 0.5, 0.33, and 0.25 years, \textit{i.e.} the Jose, Gleissberg, Hale and Schwabe cycles, the  annual cycle and its first three harmonics are identified.}

\keywords{Global Sea Level Pressure, Taylow-Couette flow, Triskel pattern}
\titlerunning{Triskels and Symmetries of Mean Global Sea-Level Pressure}
\maketitle

\section{Introduction} 
Understanding the mean circulation of air masses is one of the most ancient problems in meteorology (\textit{cf.} \cite{Lorentz1967,Lindzen1988}). Solar insolation drives the first order structure and motions of atmospheric masses. Early views thought them to be organized in planetary scale circulation cells, arranged roughly symmetrically with respect to the equator. In the troposphere, warm winds blow from the equator, forming the top of the Hadley cells (\textit{cf.} \cite{Hadley1735}); air cools, becomes denser and sinks near 30° (N and S), generating two belts of high pressures. The circuit is closed by return winds blowing towards the (low pressure) equator. Cold winds in the lower atmosphere blow from the poles to warmer regions, their density decreases as they approach 60° (N and S) latitude, where they rise. The circuit is closed by return winds in the troposphere, back to the polar high pressures, forming the polar cells. The Ferrel cells (\textit{cf.} \cite{Ferrel1856}) located between the polar and Hadley cells, extend between 30° and 60° (N and S).\\

\cite{Lindzen1988} summarize the situation at the end of the first half of the 20th century: “\textit{By the early part of this century \citep{Jeffreys1926}, the idea was being put forth that the zonally averaged circulation might, in large measure, be forced by eddies}” \ldots “\textit{\cite{Starr1948} was going so far as to suggest that the symmetric circulation was inconsequential}”. \cite{Schneider1977} and \cite{Schneider1977a} were probably the first to propose an explicit calculation of the symmetric circulation; they showed that purely symmetric circulations could maintain strong subtropical jets and contribute to the maintenance of surface winds. The symmetry of the mean heat received by Earth was sufficient to force the symmetry of the cell structure, the symmetry of the subtropical currents and the general wind pattern.\\

Based on a decade of observations (1963-1973), \cite{Lindzen1988} found that as soon as the peak heating is a few degrees in latitude off the equator, profound asymmetries in the Hadley circulation result, with the summer cell becoming negligible. The annually averaged Hadley circulation is much larger than that forced by the annually averaged heating.\\

We know that the heat coming from the Sun’s activity varies with time (\textit{e.g.} \cite{Usoskin2003,Solanki2004,Lockwood2011,Vieira2011,Abreu2012,Kutiev2013,Thuillier2014,LeMouel2020a,LeMouel2020b,Courtillot2021}) and so does the inclination of our planet’s rotation axis (\textit{e.g.} \cite{Barnes1983,Rochester1984,Lambeck2005,Schindelegger2013,Lopes2017,Barkin2019,LeMouel2019a,Krylov2020,LeMouel2021a,Lopes2021}). Consequences of these two variabilities are expected to affect the atmosphere, and in particular the climate (\textit{e.g.} \cite{Morth1979,Barnes1983,Lindzen1994,Solanki2004,Gray2010,Gray2013,Roy2010,Abreu2012,Kutiev2013,Schindelegger2013,Johnstone2014,LeMouel2019b,Gruzdev2020,Cionco2021, Connolly2021, Drews2021,Sonechkin2021}). To these already complex interactions, one must add those due to the ocean, the largest heat exchanger that warms and cools on its own (longer) time scales. \cite{Lindzen1978} was the first to show how (to first order) atmospheric circulation was forced by the physics of the ocean surface. \\

Interest in recent climate warming has led to question how the convection cells, and in particular the Hadley cells, being the main ones, react to a temperature increase (\textit{e.g.} \cite{Chang1995, Dima2003,Frierson2007,Hu2007,Kharin2007,Lu2007,Tandon2013,Shepherd2014,Tao2016,Grise2020}). Also,  how do these variations affect extreme meteorological events (\textit{e.g.} \cite{Schaeffer2005,Stott2010,Rahmstorf2011,Rummukainen2012,Trenberth2015})? Most  authors who have studied the space-time evolution of convection cells call upon the analysis in terms of components of sea-level pressure (\textbf{SLP}). \textbf{SLP} is directly related to climate indices (\textit{e.g.} \cite{Vaideanu2020,LeMouel2019b}). Decomposition methods often use “\textit{Empirical  Orthogonal Functions}” (\textbf{EOF}). In this paper, we select an alternate method: the time  decomposition of \textbf{SLP} using “\textit{Singular Spectrum Analysis}” (\textbf{SSA}, \textit{e.g.} \cite{Golyandina2013}, a method that we have used extensively with success in some recent works  (\textit{e.g.} \cite{Lopes2017,LeMouel2019a,LeMouel2020a,LeMouel2020b,Courtillot2021,LeMouel2021a,Lopes2021} \\
 
The \textbf{SLP} data we use are described in section 2, their analysis using \textbf{SSA} in section 3 and the results are discussed in section 4.

\section{The SLP Data}
\begin{figure}
    \centerline{\includegraphics[width=\columnwidth]{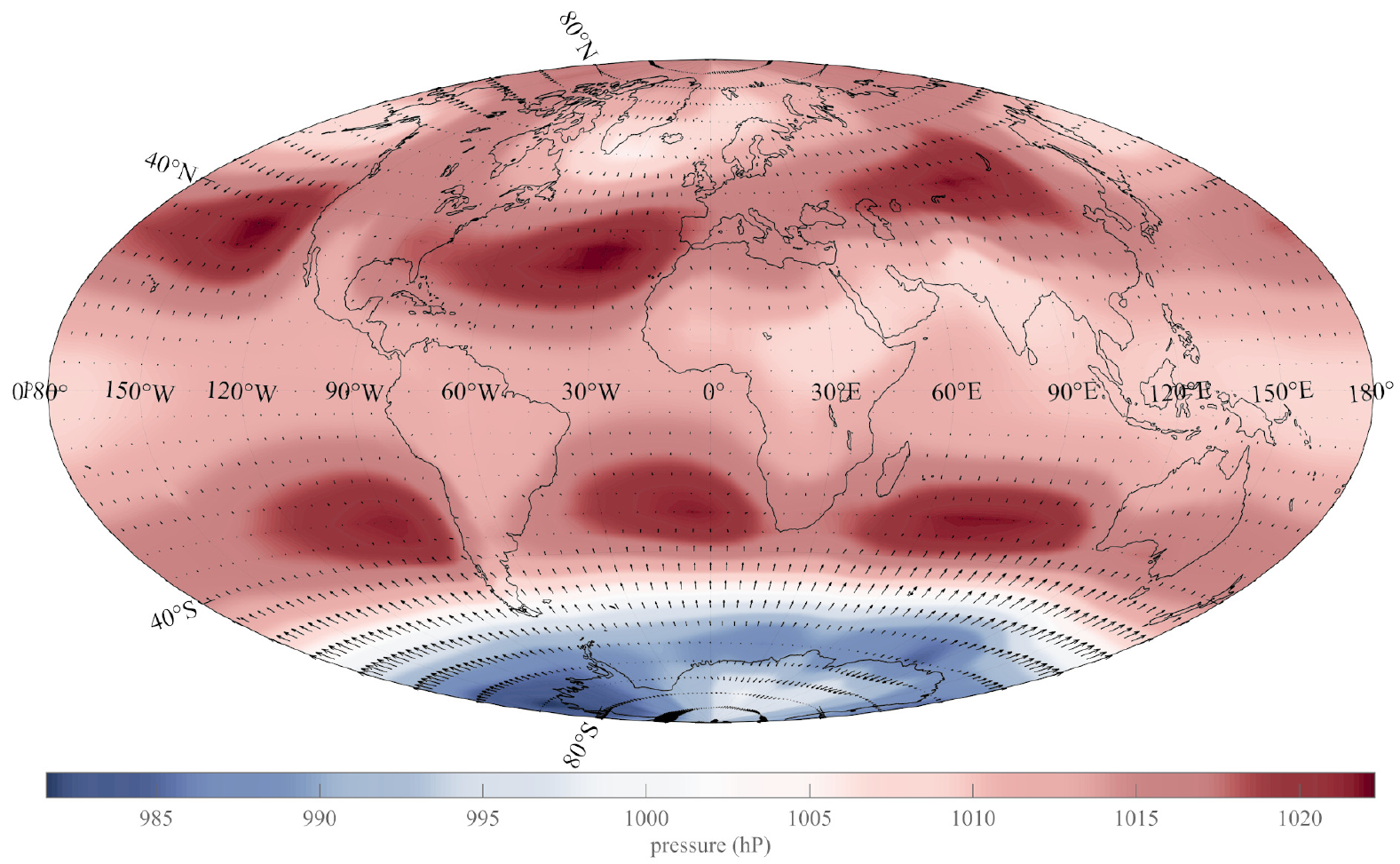}} 	
    \caption{1850 to 2020 mean pressure at sea level (\textbf{SLP}) (Hammer-Aitoff projection). }
\label{Fig:01}
\end{figure}

The \textbf{SLP} data are maintained by the Met Office Hadley Centre\footnote{https://www.metoffice.gov.uk/hadobs/hadslp2/data/download.html}. They are available in map form for global pressure every month since 1850 to the present, with a spatial sampling of 5°x5°. As indicated by the Met Office, the series is not homogeneous in time (two intervals from 1850 to 2004 and 2005 to 2020 in which the means are homogeneous but the variances differ). \cite{Allan2006} discuss the locations and number of ground observations they use to build the series. We show the map of 1850 to 2020 mean pressures in Figure \ref{Fig:01}.\\

The map of Figure \ref{Fig:01} is very similar to that obtained from simply averaging monthly \textbf{SLP} maps, demonstrating the very stable geometry of the global atmospheric structure. Well-known patterns are clearly rendered by the map: in the Southern hemisphere, the structure is dominated  by three large positive features in the southern parts of the main oceans (Pacific, Atlantic and  Indian) separated by the southern ends of the three southern continents (South America, South Africa and Australia). This approximate three-fold symmetry actually seems to “leak” into a four-fold symmetry: the positive feature extending from Australia to western South America can be described as exhibiting two weaker features, the one over Australia and East of it being the weakest.\\

The pattern in the Northern hemisphere is similar to that in the Southern hemisphere, with two features extending over the northern Pacific and Atlantic oceans, similar to the southern structure, but with the third positive anomaly lying over the Asian continents (central Asia and Tibet). This overall structure of global positive pressure features could be roughly described as the intersection of a series of three strong and one weaker cylinder, parallel to the Earth’s axis of rotation, with the earth’s surface. South of 40°S latitude, the features turn negative and form the quasi-circular Antarctic oscillation (AAO), also known as the Southern annular mode (\textbf{SAM}), a belt of low pressures surrounding the frozen continent (\textit{e.g.} \cite{Marshal2003,Gillett2006}). A zonal average of Figure \ref{Fig:01} is not really consistent with the classical latitudinal description of the Hadley, Ferrel and polar cells, with relatively moderate low pressures over the equatorial belt, high pressures around 30°N and lower values at higher latitudes, particularly over Antarctica (see also Figure \ref{Fig:04}). The overall averaged zonal structure is not symmetrical with respect to the
equator \citep{Lindzen1988}. \\

\begin{figure}[h!]
    \centerline{\includegraphics[width=\columnwidth]{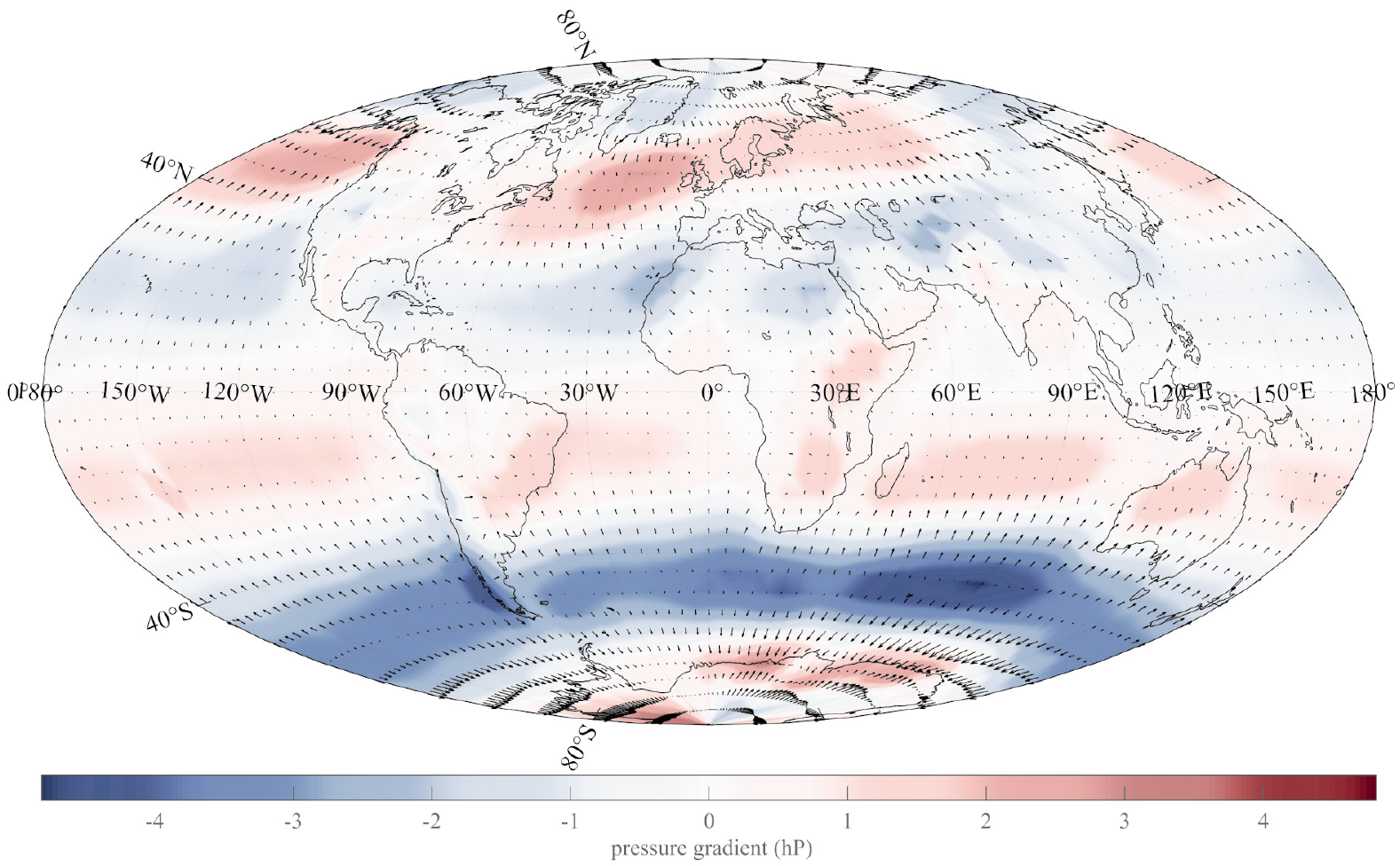}} 	
    \caption{1850 to 2020 time average of the spatial pressure gradients (i.e. winds up to a constant) at sea level (Hammer-Aitoff projection). }
\label{Fig:02}
\end{figure}

Winds are proportional to the space derivative of pressure (Figure \ref{Fig:02}, which is the map most people are familiar with), and thus in space phase quadrature with pressure features. The space derivative is a high-pass operator which exacerbates discontinuities and promotes oscillations one order higher than the causative features. Figure \ref{Fig:02} shows how winds (the gradients of pressure) are slowed or suppressed by continental masses; also, the red and blue " anomalies " are often paired, which reminds one of Laplace’s statement that the vector sum of winds must be close to zero at any instant.

\section{The SSA of SLP Data}
Rather than building a map of spatial likelihood of pressure structures at a given time, we build the time series of pressure for each couple of (longitude, latitude) coordinates. These are analyzed using the method of Singular spectrum analysis (\textbf{iSSA} , \textit{e.g.} \cite{Golyandina2013}): each time series is decomposed into a sum of a trend and periodic or quasi- periodic components. \\

Note: the trend is the first component “extracted” by \textbf{SSA}, it is a sort of mean with respect to time, a non-oscillating function of time; it should not be confused with the spatial mean of maps in time (see for example Figure \ref{Fig:02}). \\

The trend ($\sim$1008 hP), that is the first and largest \textbf{SSA} component, represents more than 70\% of the total variance (\textbf{sv}) of the original series. The sequence of the next quasi-periodic components is, in decreasing order of periods, $\sim$130 years ($\sim$0.7 hP,  $\sim$0.06\% of the \textbf{sv}; \cite{Jose1965}), $\sim$90 yr ($\sim$21 hP, $\sim$1.9\% of the \textbf{sv};  \cite{Gleissberg1939, LeMouel2017}), $\sim$50 yr ($\sim$0.2 hP, $\sim$0.02\% of the sv), $\sim$22 yr ($\sim$0.50 hP, $\sim$0.04\% of the \textbf{sv}; Hale cycle, \cite{Usoskin2017}), $\sim$15 yr ( $\sim$0.2 hP, $\sim$0.02\% of the \textbf{sv}, upper bound of the Schwabe cycle; \cite{Schwabe1844}, $\sim$4 yr ($\sim$0.3 hP, $\sim$0.03\% of the \textbf{sv}) , $\sim$1.8 yr ($\sim$0.3 hP, $\sim$0.03\% of the \textbf{sv}), then 1 yr ( $\sim$93 hP, $\sim$8.3\% of the \textbf{sv}), 0.5 yr ($\sim$65 hP, $\sim$5.8\% of the \textbf{sv}), 0.33 yr ($\sim$44 hP, $\sim$3.9\% of the \textbf{sv}) and 0.25 yr ($\sim$21 hP, $\sim$ 1.9\% of the \textbf{sv}). \\

\begin{figure}[h!]
    \centerline{\includegraphics[width=\columnwidth]{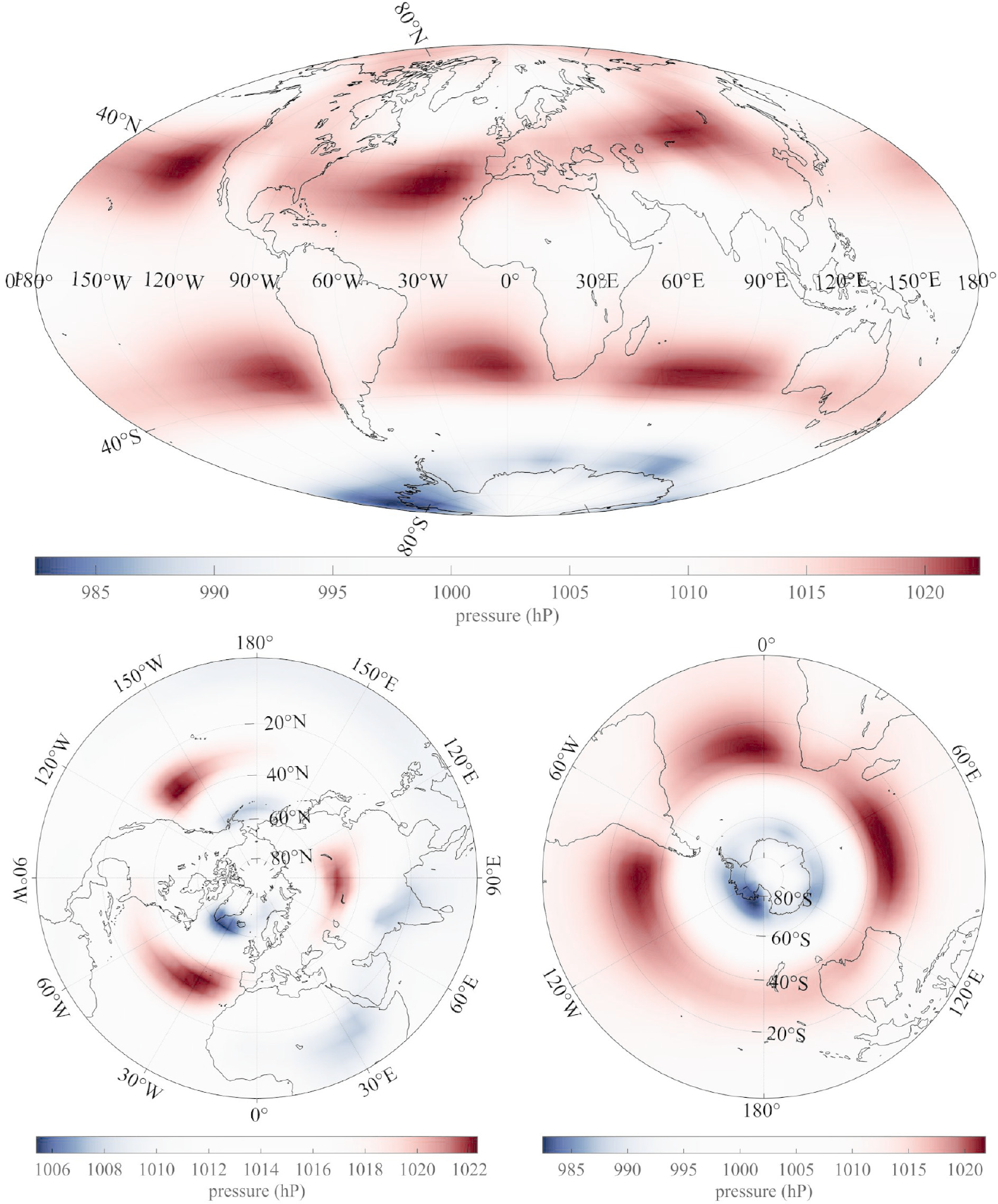}} 	
    \caption{1850 to 2020 mean of trends of pressure at sea level (\textbf{SLP}) from, determined  using Singular spectrum analysis (\textbf{SSA}) (top: Hammer-Aitoff projection; lower left: stereographic  projection of the \textbf{NH}; lower right: stereographic projection of the \textbf{SH}). }
\label{Fig:03}
\end{figure}

\begin{figure}[h!]
    \centerline{\includegraphics[width=\columnwidth]{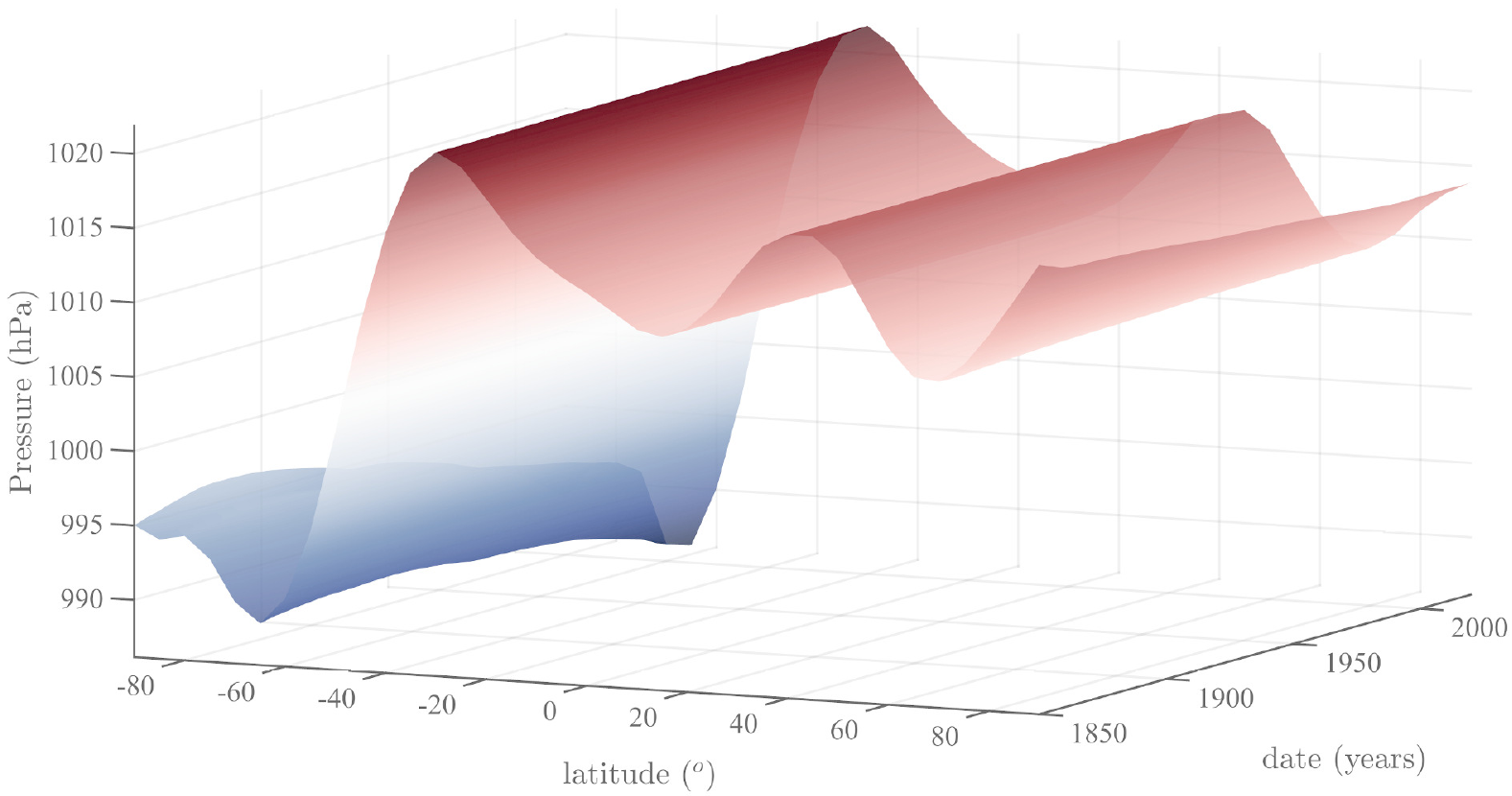}} 	
    \caption{Evolution of the \textbf{SSA} trend of \textbf{SLP} as a function of time (170 yr since 1850) and  latitude, illustrating the stability of the zonal pressure structure in time.}
\label{Fig:04}
\end{figure}

These are readily recognized as the Jose, Gleissberg, Hale and Schwabe cycles, the quasi biennal oscillation and the annual cycle followed by its first three harmonics. Figure \ref{Fig:03} represents the mean (from 1850 to 2020) of the \textbf{SSA} trends of \textbf{SLP} in Hammer-Aitoff and North and South stereographic polar projections. This mean is representative, since the observed (decreasing) overall variation of pressure is only 0.1\% in the 170 years of the study. We illustrate the stability of the \textbf{SLP} trend both in space and time with Figure \ref{Fig:04}, which shows the variation in the trend of \textbf{SLP} since 1850 as a function of latitude. Indeed, the latitudinal structure of mean pressure (\textbf{SSA} trend) is remarkably stable over 170 years (a rather rare feature in geophysics and geodynamics, both internal and external to the solid Earth), with its typical asymmetry between the two hemispheres. \\

The stereographic projections of Figure \ref{Fig:03} give a clear image of the large scale atmospheric circulation. They are close to the original mean of Figure \ref{Fig:01}, and the same large scale features appear, with sharper contours. The polar projections are particularly revealing. There is a strong negative feature south of Greenland. In the northern hemisphere, three sharp positive features lie on a circle at 40°N latitude and are located at the three apices of an equilateral triangle. The center of this 3-fold symmetry (that we will call a “triskel”, a Celtic symbol) is located near 77°N, 90°W (13° away from the North pole). In the southern hemisphere, the pattern is closer to 4-fold symmetry, with one of the four features much weaker than the other three.\\

\section{Discussion}
The physics that is appropriate to analyze the motions of fluid masses in the atmosphere, ocean (and mantle) is that of stationary turbulent flow (\textit{e.g.} \cite{Chandrasekhar1961,Frisch1995}). The large Hadley, Ferrel and polar cells are generally interpreted in terms of Taylor-Couette flow. Analytical solutions are available in the case of a cylindrical geometry (\textit{e.g.} \cite{Taylor1923}), but raise problems in the spherical case (\textit{e.g.} \cite{Schrauf1986,Mamun1995,Nakabayashi1995,Hollerbach2006,Malhoul2016,Garcia2019,Mannix2021}). \\

\cite{Forbes2018} show that a simple solution for Taylor-Couette flow of a viscous fluid between two concentric cylinders with radii $a$ and $b$ rotating in opposite directions and submitted to an azimuthal perturbation with angle $\theta$, can be written as:

\begin{subequations}
\begin{equation}
		r(\xi) = r_n + \varepsilon a \cos (q\xi) 
	\label{eq:01a}
\end{equation}
\begin{equation}
    \theta(\xi,t) = \xi - (b^2+a^2)* \omega * t* [1-\dfrac{r_n^2}{r^2(\xi)}] / (b^2-a^2)
	\label{eq:01b}
\end{equation}
\end{subequations}

in which $r_n$ is the neutral radius and $q$ is an integer corresponding to the flow mode. One has:
\begin{equation*}
    r_n = \sqrt{\dfrac{2a^2*b^2}{a^2+b^2}}
\end{equation*}

\begin{figure}[h!]
    \centerline{\includegraphics[width=\columnwidth]{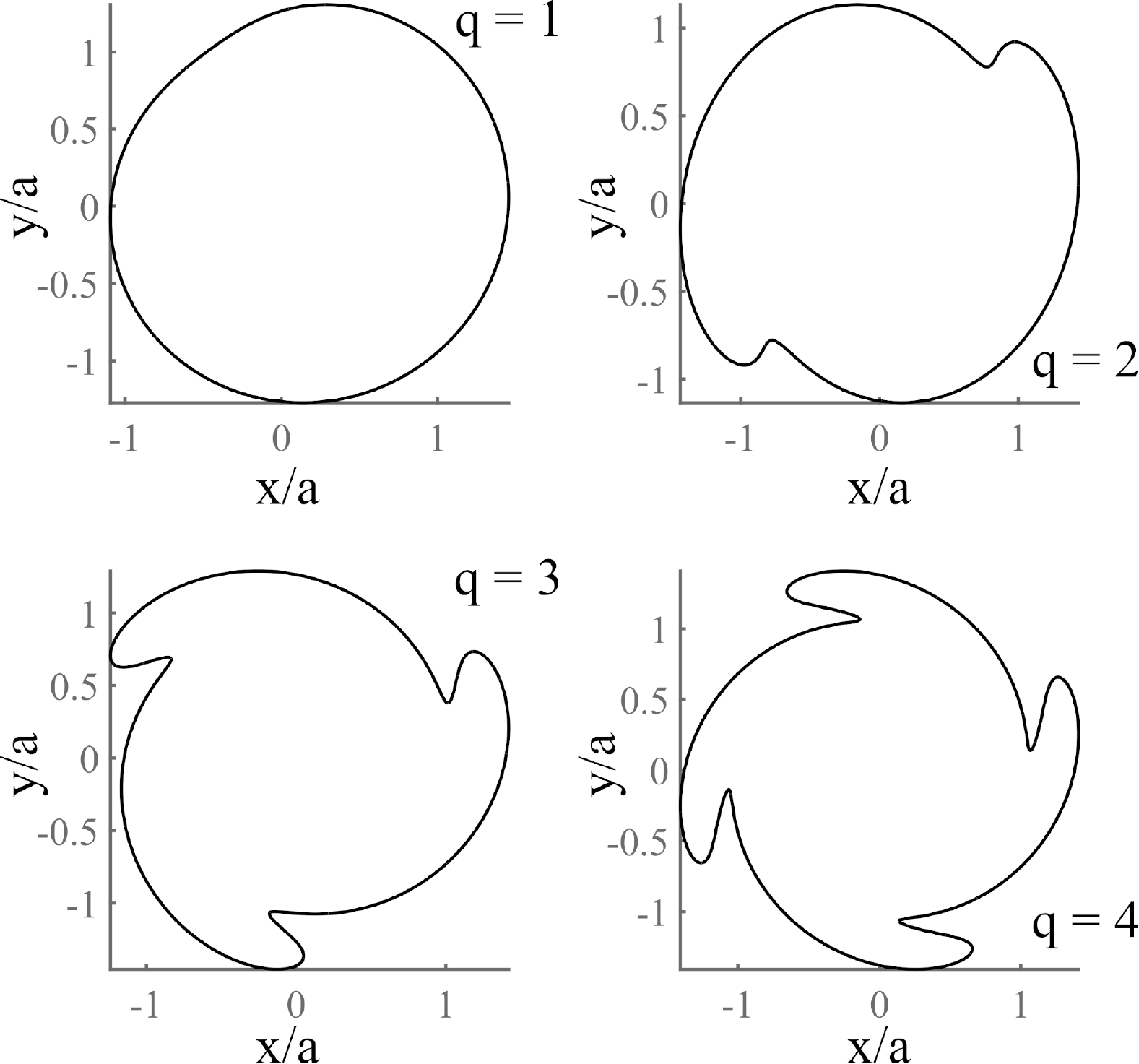}} 	 
    \caption{Solutions of system of equations  (\ref{eq:01a},\ref{eq:01b}),  for integer mode values q = 1 to 4 (see text).}
\label{Fig:05}
\end{figure}

The solutions of system (\ref{eq:01a},\ref{eq:01b}) are displayed in Figure \ref{Fig:05} for stationary flows with integers q = 1 to 4, with a = 0.5, b = 1, $\omega*t$ = 1 and $\varepsilon$ = 0.1. Modes corresponding to parameter $q$ evolve and $q$ is fixed by the nature and symmetry of the flow.\\

\begin{figure}[h!]
	\begin{subfigure}[b]{\columnwidth}
		\centerline{\includegraphics[width=\columnwidth]{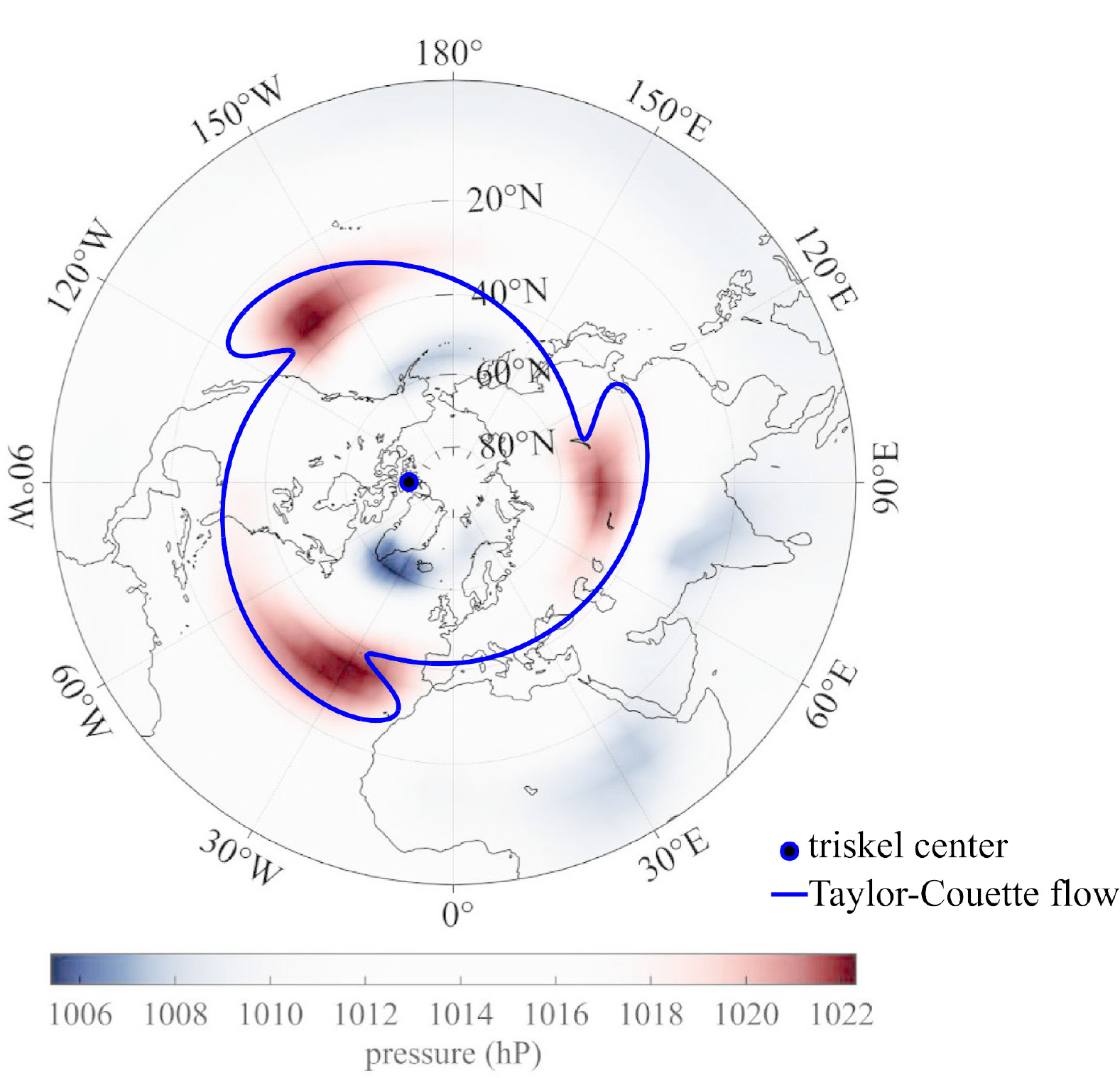}} 	
		\caption{Superimposition of the triskel Taylor-Couette pattern of flow for mode $q$ = 3 and observed mean pattern of sea-level pressure in the North hemisphere since 1850}
		\label{Fig:06a}
	\end{subfigure}
	\begin{subfigure}[b]{\columnwidth}
		\centerline{\includegraphics[width=\columnwidth]{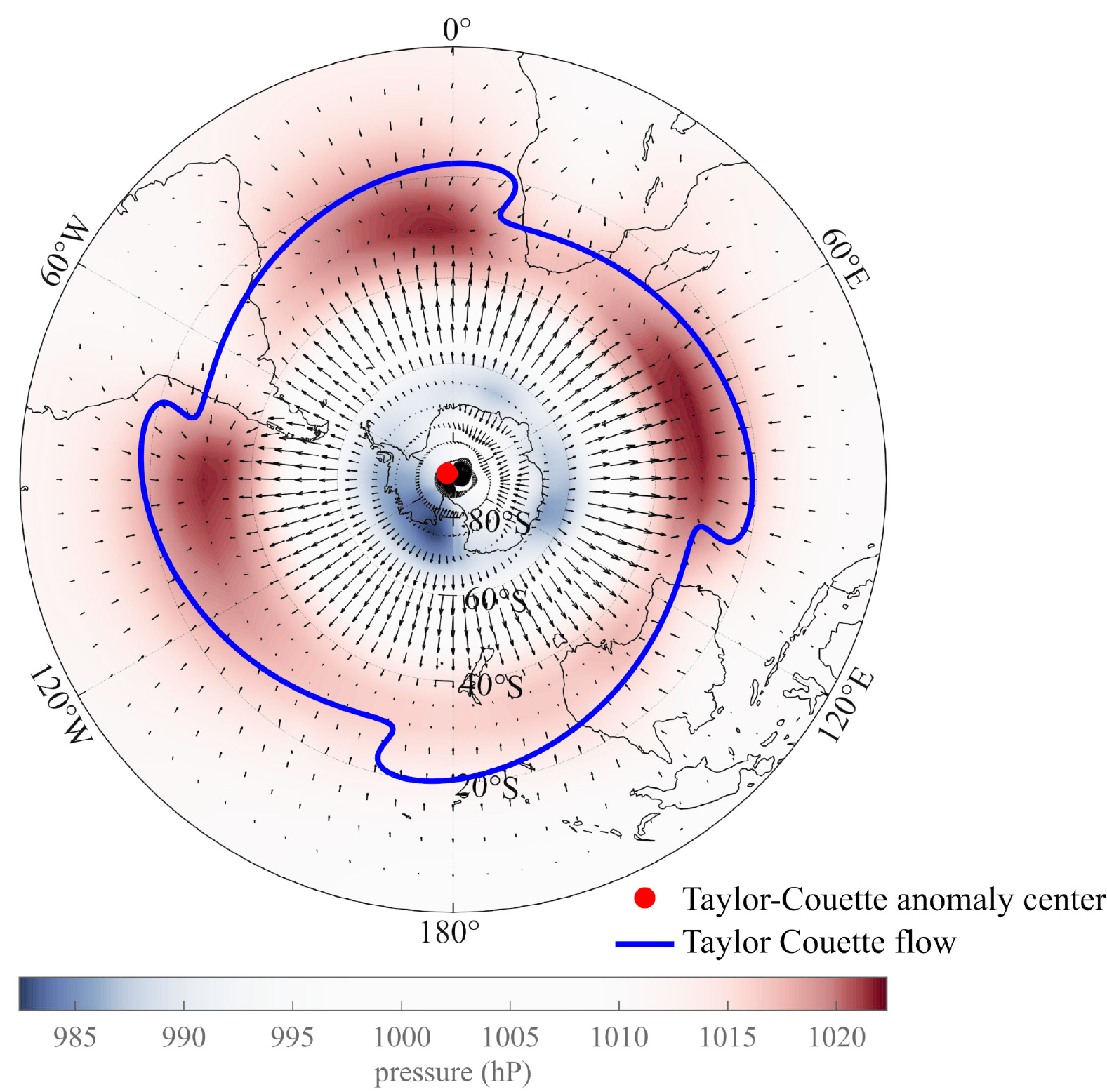}} 	
		\caption{Superimposition of the Taylor-Couette pattern of flow for mode $q$ = 4 and observed  mean pattern of sea-level pressure in the South hemisphere since 1850.}
		\label{Fig:06b}
	\end{subfigure}
	\caption{Superimposition of the Taylor-Couette patterns both North and South hemispheres.}
\end{figure}

Figure \ref{Fig:06a} shows the superimposition of a pattern of Taylor-Couette flow with mode $q$ = 3 and the \textbf{SLP} trend map of Figure \ref{Fig:03} (lower left) for the northern hemisphere. We invert equations (\ref{eq:01a},\ref{eq:01b}) by simulated annealing \citep{Kirkpatrick1983} for $\omega*t$ = 1 (flow stationary and constant in time). The amplitude of the flow perturbation $\varepsilon$ is on the order of 0.100 $\pm$ 0.002, and the triskel pattern is centered on 91.3 $\pm$ 0.1°W - 77.1 ± 0.2°N. The fit is excellent and compatible with the shearing due to the direction of winds (East to West) at mid latitudes, opposite to the direction of Earth’s rotation from West to East.\\

Figure \ref{Fig:06b} shows the superimposition of a pattern of Taylor-Couette flow with mode $q$ = 4 and the \textbf{SLP} trend map of Figure \ref{Fig:03} (lower right) for the southern hemisphere. The amplitude of the flow perturbation $\varepsilon$ is reduced to 0.05 $\pm$ 0.01, and the symmetry 4 is centered on 88.1 $\pm$ 0.2°W  - 7.5 $\pm$ 0.1°N. \\

The Hadley, Ferrell or polar type cells change with season. This is why the zonal structure of the atmosphere is shown in seasonal (monthly, bi-monthly) maps that can be averaged over several years. These oscillations cannot be analyzed through spatial decomposition methods such as \textbf{EOF} or spherical harmonics. On the other hand, SSA is a scalar time analysis and can extract from the data temporal features. We have shown that atmospheric surface pressures display simple geometric/geographic features that are stable both in space and time. The oscillatory components
belong to the harmonic families of the Schwabe (33, 22, 11, 5.5, 3.3 and 1.6 yr) and annual (1, 1/2, 1/3, 1/4 and 1/5 yr) cycles, plus the dominant 90 yr Gleissberg cycle. \\

Finally, one can only be struck by the kinship between the axial cylinders that form the main components of the sea-level pressure structure (with the quasi-symmetry of the outline of these cylinders, yet the failure of perfect symmetry, with 3 cylinders in the northern hemisphere and more like 4 in the southern hemisphere). These are geometrically reminiscent of the axial cylinders found by \cite{Gubbins1987} and \cite{Gubbins1993} in their analyses of flow in the Earth’s molten outer core that generates the geomagnetic field. The physics in both cases appears quite distinct but the analogy may indicate a path to finding the mechanism that generates these features.

\section{Conclusion}
In this note, we have analyzed the time evolution of mean sea-level pressure \textbf{SLP} since 1850, using the \textbf{iSSA} method of spectral analysis. The method decomposes the original time series into a set of time dependent components and a trend. The quasi-periodic components are in decreasing order of periods, cycles of approximately 130, 90, 50, 22, 15, 4, 1.8, 1, 0.5, 0.33, and 0.25 years. These are recognized as the Jose, Gleissberg, Hale and Schwabe “long” (decadal and  multi-decadal) cycles (with some harmonics of the well-known Schwabe “solar” cycle\footnote{The Schwabe quasi-cycle is traditionally given as being 11 yr long but actually spans the 9 to 14 yr period range.} ) and the “short” annual cycle and its first three harmonics. These periods are encountered in many solar and terrestrial phenomena and are now attributed to commensurable periods of the Jovian planets  \citep{Morth1979,Lopes2021}. \\

As far as the general trends of sea-level pressure extracted with SSA are concerned, they have fluctuated by less than 0.1\% in 170 years (or 6.10 -3 hP/yr). The mid latitudes of each  hemisphere of the Earth’s surface harbor a set of positive features (on the order of 20 hP), all but one (over central Asia and Tibet) lying over the main ocean basins (Figure \ref{Fig:03}, top). This is in full  agreement with \cite{Schneider1977}, who see the oceans as the main engine driving the 194 general circulation of the atmosphere. In stereographic polar projection, the northern hemisphere displays a set of three positive features, forming an almost perfect equilateral triangle or triskel  (Figure 03, lower left). The southern hemisphere also features a set of three main positive features but arranged as an isosceles triangle, with a possible fourth (but much fainter) feature (Figure \ref{Fig:03}, lower right). A preliminary analysis suggests that the atmosphere in the two hemispheres could be the site of Taylor-Couette (sheared) differential flow of mode 3 (N hemisphere) or mode 4 (S hemisphere). The mode is imposed by the geometry of the oceanic vs continental nature of the Earth’s surface. The remarkable regularity and order three symmetry of the northern hemisphere triskel occurs despite the lack of cylindrical symmetry of the northern continents. The stronger intensity and larger size of features in the southern hemisphere is due to the presence of the annular flow of cold air over the southern ocean and the obstacles due to the southern parts or tips of South America, Southern Africa and Australia.\\

Singular spectrum analysis is a scalar analysis as a function of time, which is useful in detecting time varying patterns and features. Spatial components that \textbf{SSA} extracts are not artefacts, contrary to what could happen when using spherical harmonics or \textbf{EOF}. A component is present only if there is spatial coherency in the observational data. In the case of sea-level pressure, we find stable (non oscillatory) components/structures/features (as we have called them), The stability of pressure trends over the 170 years of the analysis implies that the idea of eddies maintaining the cells and forcing their oscillations is no more warranted. Taylor-Couette flow between rotating cylinders provides a good fit to the observations. The geometrical analogy with the cylinders generating the geodynamo may be a promising path to solving the dynamics of the large scale circulation in the Earth’s atmosphere.

\newpage
\bibliographystyle{aa}

\end{document}